\documentclass[prl,twocolumn]{revtex4}
\usepackage{graphicx}

\begin{document}
\draft
\preprint{}
\title{Evidence for Thermally Activated Spontaneous Fluxoid Formation in
Superconducting Thin-Film Rings}
\author{J. R. Kirtley and C. C. Tsuei}
\affiliation{ IBM T.J. Watson Research Center, P.O. Box 218, Yorktown Heights, NY 10598}
\author{F. Tafuri}
\affiliation{INFM-Coherentia Dip. di Ingegneria dell'Informazione, Seconda
Universit\'{a} di Napoli, Aversa (CE), Italy}
\date{\today}
\begin{abstract}
We have observed spontaneous fluxoid generation in thin-film rings of the
amorphous superconductor Mo$_3$Si, cooled through the normal-superconducting
transition, as a function of quench rate and externally
applied magnetic field, using a variable sample temperature scanning
SQUID microscope. Our results can be explained using a model of freezout
of thermally activated fluxoids, mediated by
the transport of bulk vortices across the ring walls.
This mechanism is complementary to a mechanism
proposed by Kibble and Zurek, which only relies on
causality to produce a freezout of order parameter fluctuations.
\end{abstract}
\maketitle
%\begin{multicols}{2}
%\narrowtext
%\pacs{73.40.Gk,73.40.Rw,74.50.+r,74.72.-h}

\pagebreak
%\narrowtext
A number of years ago Zurek \cite{zurek} proposed that condensed matter
systems can be used as models for testing ideas,
first introduced by Kibble\cite{kibble,kibble2}, concerning the
quenching of topological defects in the early evolution of the universe.
The principal picture developed by Kibble and Zurek was that, as a system approaches
the critical temperature of a phase transition from above, the
system slows down until at some temperature different regions
cannot communicate with each other quickly enough to have correlated
order parameters. As the transition proceeds
further, fluctuations between different regions are ``frozen in", leading
to relatively long-lived topological defects. Although this was only
one of several mechanisms for freezout of topological defects considered by
Kibble and Zurek, for brevity we will refer to this as the ``Kibble-Zurek"
mechanism in this paper.
The advantage of using
condensed matter systems to test these ideas is clear: while it is
impossible to do experiments on a cosmological scale,
there are a
number of well understood phase transitions into macroscopic quantum states
in condensed matter
that are experimentally accessible.

The results of a number of experimental tests of these ideas have
been mixed: Although vortex
generation
was at first observed in $^4$He upon rapid quenching into the superfluid
state \cite{hendry}, a refinement of the original experiment by the same
group \cite{dodd} found a residual vortex density two orders of magnitude
smaller than the Kibble-Zurek prediction. Experiments on vortex formation
upon rapid quenching in $^3$He \cite{bauerle,ruutu} found good agreement
with the Kibble-Zurek predictions. A search for spontaneous vortex generation
in a bulk superconductor upon rapid quenching \cite{carmi} was unsuccessful,
although sensitivity of 10$^{-3}$ of the Kibble-Zurek prediction was reported.
However, tests on superconducting loops interrupted by Josephson weak links
have found good agreement with the Kibble-Zurek predictions, both for
high-T$_c$ \cite{carmi2} and low-T$_c$ \cite{kavoussanaki,monaco,monaco2}
superconductors.

In addition to the standard Kibble-Zurek mechanism, which only relies on
the satisfaction of the principle of causality to ``freeze in" topological defects,
it is also possible for
thermally activated defects to appear and be trapped in the
system during cooldown \cite{zurek}. Detailed calculations of trapping
of thermally activated fluctuations
have been made in the hot Abelian Higgs model\cite{hindmarsh}, and for
the case of superconducting rings\cite{ghinovker}. In this paper we report
on measurements of spontaneous fluxoid formation in arrays of thin film superconducting
rings with various ring dimensions
and compositions. SQUID imaging of such arrays allowed the performance
of a number of cooling experiments simultaneously. The film thicknesses and
compositions were chosen to obtain large magnetic penetration depths
and inductances, which produce favorable conditions for
thermally activated behavior. The experiments provide evidence
for the thermally activated
appearance of topological defects (in this case vortices and
antivortices) in systems undergoing a sufficiently rapid
phase transition into an ordered state. We argue
that such processes should be considered as a complementary mechanism
in tests of the Kibble-Zurek predictions.

Although we obtained qualitatively similar results on rings fabricated
from the low-T$_c$ superconductor Nb, and the cuprate superconductor
YBa$_2$Cu$_3$O$_{7-\delta}$, we will report in this paper on
thin-film ($d$=50nm) rings of
the amorphous superconductor Mo$_3$Si\cite{tsueirev}, which showed the
largest spontaneous fluxoid generation, and on which the
most detailed measurements were taken. The films were
fabricated by rf-sputtering on sapphire substrates
and photolithographically defined using
ion-beam etching. Results will be presented here on two sets of
rings. One set consisted of 25 circular rings, with
inside diameter 2$a$=40 $\mu$m and  outside diameter 2$b$=80 $\mu$m,
spaced by 160 $\mu$m in a 5$\times$5 square array.
The second set had 144 rings with 2$a$=20 $\mu$m and 2$b$=30 $\mu$m,
spaced by 60 $\mu$m in a
12$\times$12 square array.
The films had a resistive superconducting critical temperature
T$_c$=7.81 K, with a transition
width (10\%-90\%) $\Delta T_c$=0.03 K wide. Magnetic
imaging measurements were made with a variable sample temperature, scanning
SQUID microscope\cite{vartapl},
with a square pickup loop 8 $\mu$m on a side. The sample was mounted with
a thin layer of GE varnish on a saphire substrate to which a silicon diode
thermometer and non-inductive resistive heater were attached. Both sample and SQUID
were inside an evacuated can surrounded by liquid $^4$He. The thermal time
constant of the sample mount was varied by changing the amount of $^3$He exchange gas
between it and the $^4$He bath.
%Fits of SQUID microscope images of
%Pearl vortices \cite{pearl,vartapl} in
%the Mo$_3$Si films before etching gave a London penetration depth
%$\lambda_L$=0.7$ \mu$m at T=4.4 K, corresponding to a Pearl length of
%$\Lambda$=2$\lambda_L^2$/d=20$\mu$m.
Fits to SQUID images of the rings
with fluxoid number $N$=1 at zero field gave a Pearl length \cite{pearl}
$\Lambda$=2$\lambda_L^2$/d ($\lambda_L$ the London penetration depth)
=7 $\mu$m at 4.4 K.

Our experiments consisted of cooling the
sample in a given field at a controlled rate, and then imaging the rings with
the SQUID microscope to determine the final fluxoid state.
The spacing between
rings was large enough that ring-ring interactions were not important in the
cooldown dynamics.

\begin{figure}
\includegraphics[width=3.5in]{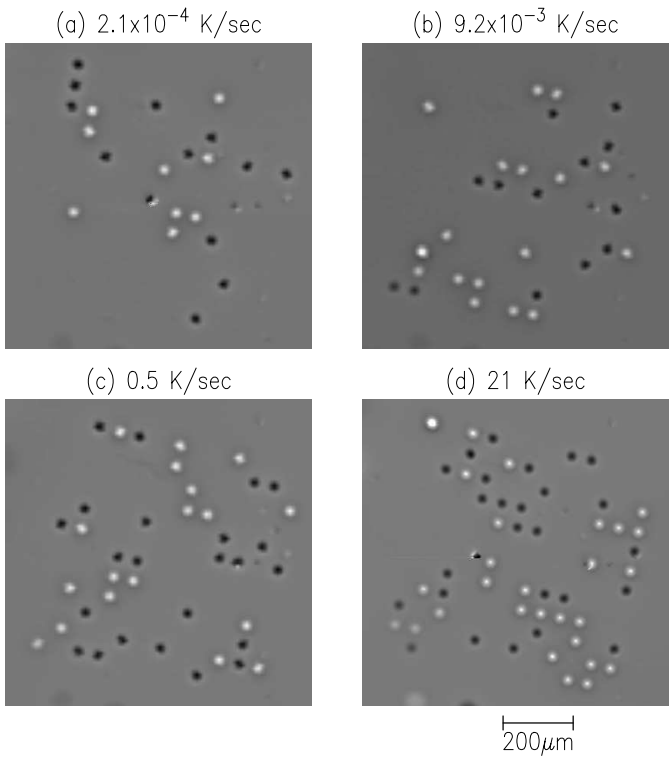}
\vspace{0.1in}
\caption{\label{fig:cosmprl1}Scanning SQUID microscope images of a
12x12 array of 20 $\mu$m inside diameter, 30 $\mu$m outside diameter
thin film rings of Mo$_3$Si, cooled in zero field through the superconducting T$_c$
at various rates, and imaged with
an 8 $\mu$m square pickup loop. The full scale variation of the grey scale
images corresponded to 0.25$\Phi_0$ (a), 0.3$\Phi_0$ (b), 0.25 $\Phi_0$ (c), and
0.37$\Phi_0$ (d) change in flux through the pickup loop. The principal axes
of the square array are rotated by 12$^o$ clockwise with respect to the image frame.}
\end{figure}

Figure 1 shows SQUID microscope images of the 20 $\mu$m/30 $\mu$m ring array,
cooled through T$_c$ at zero field with varying cooling rates. In this image
the rings
with final fluxoid number $N$=0 are not visible; rings with positive
fluxoid number (counter-clockwise circulating supercurrents) are white; and those
with negative fluxoid number (clockwise supercurrents) are black.
Most of the rings with non-zero fluxoid number have N=+1 or -1,
but a few rings also have N=+2 or -2. For example,
the upper left most ring in Fig. \ref{fig:cosmprl1} (d) has N=2.

Figure 2 displays the results, plotted as the probability for a ring to be
in the N=+1 fluxoid state, for a number of cooling rates, for both ring sizes.
The error bars are assigned by setting the standard deviation of the number
of N=+1 rings to be the square root of the number of
N=+1 rings observed in each experiment. This plot shows that
P(N=+1) varies roughly linearly with the logarithm
of the cooling rate.

\begin{figure}
\includegraphics[width=3.5in]{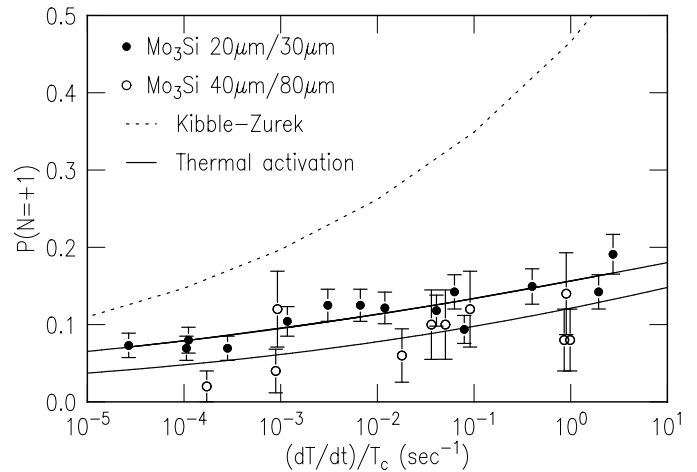}
\vspace{0.1in}
\caption{\label{fig:cosmprl3}Probability of 20 $\mu$m inside diameter, 30 $\mu$m
outside diameter Mo$_3$Si rings (closed symbols) and 40 $\mu$m/80 $\mu$m rings
(open symbols)
having a final fluxoid number of +1, as
a function of the cooling rate, after cooling in zero externally
applied magnetic flux.
The dashed line is the prediction of the Kibble-Zurek quenched
disorder model, and the
solid lines are fits to the frozen thermally activated bulk vortex model
discussed in the text.}
\end{figure}

The Kibble-Zurek prediction for the expected fluxoid number for a
quenched superconducting ring due to freezout of order parameter fluctuations
is \cite{zurek}
\begin{equation}
N=\frac{1}{2\pi}\sqrt{C/\xi_0}(\tau/\tau_Q)^{\frac{1}{8}},
\label{eq:nphi}
\end{equation}
where $C$ is the circumference of the ring, $\xi_0$ is the zero
temperature coherence length, $\tau=\pi\hbar/16k_BT_c$ is the
Ginzburg-Landau relaxation time, and $\tau_Q^{-1}= \mid dT/dt \mid/T_c$ is
the normalized cooling rate. The dashed curve in Fig. \ref{fig:cosmprl3}
is this prediction for the 20 $\mu$m/30 $\mu$m rings, using
$\xi_0$=4.9nm\cite{samoilov}, and taking $C$ to be the inside circumference
of the rings. The corresponding prediction for the
40 $\mu$m/80 $\mu$m rings would be scaled vertically by a factor of $\sqrt{2}$.
The Kibble-Zurek mechanism predicts more spontaneous fluxoid generation than
is observed, as well as a strong increase with ring size which is not observed.
Moreover, the Kibble-Zurek prediction, although it has a weak
(T$_c^{-1}\mid $dT/dt$ \mid )^{1/8}$ dependence,
nevertheless varies more rapidly with cooling rate than experiment.
We should note that the Kibble-Zurek predictions for a superconducting
ring were for ring walls much thinner than the ring diameter\cite{zurek}.
However, we do not believe that our finite ring wall widths would
effect the Kibble-Zurek predictions qualitatively.

We suggest that our observed logarithmic dependence of non-zero fluxoid
probability on cooling rate is due to spontaneous changes in the ring fluxoid
number by thermal activation in,
and transport across, the ring wall of bulk vortices.
Such a model has been used to
qualitatively explain measurements on telegraph noise
and fluxoid transition rates in superconducting thin-film rings of
underdoped cuprate Bi$_2$Sr$_2$CaCu$_2$O$_{8-\delta}$\cite{ringcond,ringprb},
which have
penetration depths and coherence lengths comparable to the present samples.
Physically our picture is that, as the rings
cool through T$_c$
thermally activated fluctuations rapidly change the fluxoid number.
The thermal activation barrier to this process becomes large
as the temperature is
reduced further, eventually turning this process off, leaving the rings
with some probability of being in energetically excited states.

We calculate the probability $P_N$ of a ring being in the Nth fluxoid number
state as follows: Consider a set of ring fluxoid states $N$, with
ring energies $E_N$, and with barrier energies $E_B$
(in our model the energy required to nucleate a vortex in the ring wall)
that must be overcome
to make the transition between them. The transition rate $\nu_{N \rightarrow N^\prime}$
is given by
\begin{equation}
\nu_{N \rightarrow N^\prime} = P_N \nu_0 e^{-(E_B-E_N)/k_BT},
\label{eq:nunnp}
\end{equation}
where $\nu_0$ is an attempt frequency, and we neglect the dependence of the
transition rate on the final state energy.
At high temperatures, in equilibrium,
$\nu_{N \rightarrow N{^\prime}}=\nu_{N^\prime \rightarrow N}$
for all $N$, $N^\prime$. This implies that
\begin{equation}
P_N=e^{-E_N/k_BT}/\sum_{N^\prime} e^{-E_{N^\prime}/k_BT}.
\label{eq:probn}
\end{equation}
The ring energies $E_N=E_r(N-\phi_a)^2$, with $\phi_a=\Phi_a/\Phi_0$ the
normalized externally applied flux, and $\Phi_0=hc/2e$ the superconducting
flux quantum\cite{barone}.
Eq. \ref{eq:probn}
implies that the probability of a ring having a non-zero fluxoid number
depends on the ratio $E_r/k_BT$. As the ring cools down, the transition
rates slow down when $E_B(T) \simeq k_BT$, fixing the ring fluxoid number.
We can assign a ``freezing" temperature T$_F$ by insisting that the
probability of there being a further fluxoid jump, obtained by
integrating the transition
rates over temperature from T=T$_F$ to T=0, be less than 1. It can then be
shown that
\begin{equation}
k_B T_F = -E_B(T_F)/\ln
\left ( \frac{\gamma \mid dT/dt \mid}{k_B\nu_0T_F} \right ),
\label{eq:chisubr}
\end{equation}
where $\gamma=dE_B(T)/dT\mid_{T=T_F}$.
We then rewrite Eq. \ref{eq:probn} to obtain the probability $P_{N,F}$
that a ring is frozen into a state with fluxoid number $N$:
\begin{equation}
P_{N,F}=e^{-\chi_r(N-\phi_a)^2}/\sum_{N^\prime} e^{-\chi_r(N^\prime-\phi_a)^2},
\label{eq:probnfrozen}
\end{equation}
with $\chi_r \equiv E_r(T_F)/k_BT_F$.
For the parameters of our system, $P_{N,F}$ varies nearly
linearly with $\chi_r$, and therefore
depends roughly logarithmically on cooling rate through Eq. \ref{eq:chisubr}.

\begin{figure}
\includegraphics[width=3.5in]{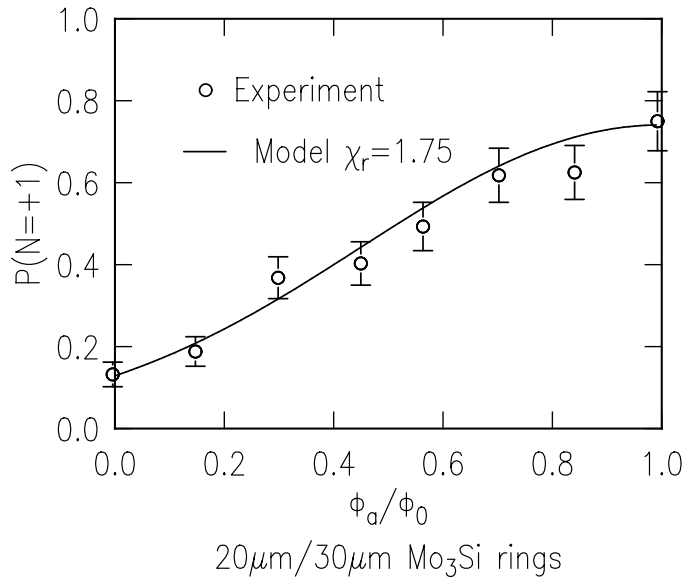}
\vspace{0.1in}
\caption{\label{fig:cosmprl2}Probability of 20 $\mu$m inside diameter, 30 $\mu$m
outside diameter Mo$_2$Si rings
having a final fluxoid number of +1, as
a function of the externally applied magnetic flux. The open symbols are the
data, the solid line is a fit to the thermally activated vortex
model discussed in the text.}
\end{figure}

The upper solid line in Fig. \ref{fig:cosmprl3} is a fit of Eq. \ref{eq:probnfrozen}
to the 20 $\mu$m/30 $\mu$m ring data, resulting in best fit values of
$E_r(T_F)/E_B(T_F)$=0.097$\pm0.1$ and
$\gamma/k_B \nu_0$=2.4$\times$10$^{-7}$ sec., with upper and lower
limits of 17.5$\times$10$^{-7}$ sec. and 0.4$\times$10$^{-7}$ sec.
respectively on the latter value,
using a doubling of $\chi^2$ as a criterion.
In the temperature range of interest,
1-t=(T$_c$-T)/T$_c \ll$ 1, the Pearl length $\Lambda(T) \simeq \Lambda(0)/(1-t^4)$
is larger than typical ring dimensions.
In this limit\cite{ringprb}
\begin{equation}
E_r(\Lambda > a,b) \simeq \Phi_0^2/(8\pi^2\Lambda).
\label{eq:er}
\end{equation}
The energy $E_B$ required to
nucleate a vortex
in a thin film strip of width $W$ is given by \cite{koganprb}
\begin{equation}
E_B= \frac{\Phi_0^2}{8\pi^2\Lambda} \ln \left ( \frac{2W}{\pi\xi} \right ) .
\label{eq:ebarrier}
\end{equation}
Combining Eq.'s \ref{eq:er} and \ref{eq:ebarrier}
we estimate $E_r/E_B \approx 1/\ln(2W/\pi\xi)$. Using Eq. (\ref{eq:ebarrier}),
Eq. (\ref{eq:chisubr}), $\Lambda$(0)=7 $\mu$m,
$\xi(T)=\xi(0)/\sqrt{1-t}$, and $W=b-a$ = 5 $\mu$m, leads to $(T_c-T_F)/T_c$=
2$\times$10$^{-3}$, $\gamma$=2.7$\times$10$^{-14}$ erg/K, leading to
$E_r/E_B$=0.30, a factor of three larger than our fit value.
Our fit value
for $\gamma/k_B \nu_0$ implies $\nu_0$=8.3$\times$10$^8$ sec.$^{-1}$,
with upper and low limits of 6.1$\times$10$^9$ sec.$^{-1}$ and
1.1$\times$10$^8$ sec.$^{-1}$ respectively, in
reasonable agreement with attempt frequencies obtained from fluxoid dynamics
of Bi$_2$Sr$_2$CaCu$_2$O$_{8-\delta}$ rings with similar penetration depths
\cite{ringcond,ringprb}. The lower solid line in Fig. \ref{fig:cosmprl3} is
a fit to the 40 $\mu$m/80 $\mu$m ring data, using the value for $\gamma/k_B\nu_0$
obtained from the 20 $\mu$m/30 $\mu$m rings, and allowing $E_r(T_F)/E_B(T_F)$ to vary.
The best fit value in this case is $E_r(T_F)/E_B(T_F)$=0.12, with
upper and lower limits of 0.15 and 0.105,
to be compared with a
calculated estimate for these rings of $E_r/E_B$=0.2. Note, however, that the
predicted dependence on wall width does not match the experimental results:
doubling the size of the ring wall should decrease the
value for $E_r/E_B$ by 30\%, but our fit values increase by about that amount.

A further test of our model is provided by measurements of the final state fluxoid
number probability as a function of applied field. Figure \ref{fig:cosmprl2}
shows the results of such experiments for the 20 $\mu$m/30 $\mu$m rings,
cooled at a cooling rate of 15 K/sec. At first these results are startling:
the fluxoid number probabilities depend only weakly on applied field. This
is in contrast to pioneering work by Davidovi\'{c} et al.
\cite{davidovic1,davidovic2} on 1.6 $\mu$m outside diameter thin film aluminum
rings, which showed, e.g.
P$_{1,F}$ stepping up sharply from 0 to 1 at $\phi_a$=0.5. This difference can
be explained qualitatively as follows: our rings, which have large values for $a,b$,
and $\Lambda$, have small ring energy spacings (scaled by $E_r$). Therefore $k_B T_F$
is comparable to $E_r(T_F)$ and it is probable that energetically unfavorable
final fluxoid numbers will be populated. On the other hand, the rings of
Davidovi\'{c} have much smaller values for $a,b$, and $\Lambda$, and the ring
energy spacings will be correspondingly larger.
The solid line in Fig. \ref{fig:cosmprl2} is a fit to the data
using Eq. \ref{eq:probnfrozen}, with $\chi_r$ as a fitting parameter.
The best fit is obtained for $\chi_r=1.75$, compared to a value of 4.3
calculated from Eq. \ref{eq:chisubr}.

In conclusion, we have found that our results for the spontaneous
generation of energetically unfavorable final fluxoid numbers in cooled
thin-film rings of the amorphous superconductor Mo$_3$Si can be explained
using a model of thermally activated bulk vortices.
This mechanism occurs over
a much different time and temperature scale than the conventional
Kibble-Zurek mechanism, which
predicts that freezout of order
parameter fluctuations occurs at $\hat{\epsilon} \equiv \mid
(T - T_c)/T_c \mid = \sqrt{\tau_0/\tau_Q} \approx 1 \times 10^{-7}$.
For our rings the thermally activated processes persist to
$\hat{\epsilon} \approx 1 \times 10^{-3}$. In this case any fluxoids generated by the
conventional Kibble-Zurek mechanism would be ``washed out" by the thermally
activated vortex mechanism. Our rings are exceptionally susceptible
to thermally activated vortex processes: they have very large magnetic
penetration depths and inductances,
which make nucleation of vortices not too energetically
costly at temperatures relatively far from T$_c$. Nevertheless, the possibility
of thermally activated processes should be considered, keeping in mind that
not only short coherence lengths, but also short penetration depths,
are desirable when using superconductors for
tests of the Kibble-Zurek mechanism for
topological defect formation in quenched macroscopic
quantum systems.

We would like to thank V.G. Kogan, and K.A. Moler for useful discussions
during the course of this work. F.T. acknowledges the support of the
ESF Project ``VORTEX".

\end{document}